\documentclass[lettersize,journal]{IEEEtran}
\usepackage{amsmath,amsfonts}
\usepackage{algorithm}
\usepackage{array}
\usepackage[caption=false,font=normalsize,labelfont=sf,textfont=sf]{subfig}
\usepackage{textcomp}
\usepackage{stfloats}
\usepackage{url}
\usepackage{verbatim}
\usepackage{graphicx}
\usepackage{cite}
\usepackage{tikz}
\usepackage{bm}
\usepackage{booktabs}
\usepackage{hyperref}
\usepackage{adjustbox}
\usepackage{xspace}
\usepackage{svg}
\usepackage{soul}
\usetikzlibrary{positioning,calc,decorations.markings,shapes,arrows,arrows.meta,backgrounds}
\usepackage{pgfplots}
\usepackage{standalone}
\pgfplotsset{compat=newest}
\usepgfplotslibrary{groupplots}
\usetikzlibrary{patterns}
\usetikzlibrary{patterns.meta}
\usetikzlibrary{external}
\usepackage{algpseudocode}
\def \qed {\hfill \vrule height6pt width 6pt depth 0pt}
\usepackage{mathrsfs}
\usepackage{xcolor}

\allowdisplaybreaks[4]

\newtheorem{lemma}{Lemma}

\newtheorem{remark}{Remark}

\DeclareMathOperator{\diagon}{diag}
\newcommand{\diag}[1]{\diagon(#1)}
\newcommand{\etal}{et al.\xspace}
\pgfmathsetmacro{\RadToDeg}{180/3.14159265359}
\newcommand{\dt}[1]{_{#1}}
\newcommand{\dti}[2]{_{#1}^{#2}}

\begin{document}

\title{An Adaptive Data-Enabled Policy Optimization Approach for Autonomous Bicycle Control}

\author{Niklas Persson, \IEEEmembership{Student member, IEEE}, Feiran Zhao, Mojtaba Kaheni, \IEEEmembership{Senior Member, IEEE}, Florian D\"orfler, \IEEEmembership{Senior Member, IEEE}, Alessandro V. Papadopoulos, \IEEEmembership{Senior Member, IEEE}
\thanks{This work was supported by the Knowledge Foundation (KKS) with grant ``Mälardalen University Automation Research Center (MARC)'', n. 20240011.}
\thanks{N. Persson, M. Kaheni, and A.V. Papadopoulos are with the Division of Intelligent Future Technologies, M\"alardalen University, 721 23 V\"aster{\aa}s, Sweden.    (e-mails: niklas.persson@mdu.se, mojtaba.kaheni@mdu.se, alessandro.papadopoulos@mdu.se).}
\thanks{F. Zhao and F. D\"orfler are with the Department of Information Technology and Electrical Engineering, ETH Zurich, 8092 Zurich, Switzerland. (e-mail: zhaofe@control.ee.ethz.ch,
dorfler@ethz.ch}
\thanks{Manuscript received February 17, 2025}}

\maketitle

\begin{abstract}
This paper presents a unified control framework that integrates a Feedback Linearization (FL) controller in the inner loop with an adaptive Data-Enabled Policy Optimization (DeePO) controller in the outer loop to balance an autonomous bicycle. While the FL controller stabilizes and partially linearizes the inherently unstable and nonlinear system, its performance is compromised by unmodeled dynamics and time-varying characteristics. To overcome these limitations, the DeePO controller is introduced to enhance adaptability and robustness.
The initial control policy of DeePO is obtained from a finite set of offline, persistently exciting input and state data. To improve stability and compensate for system nonlinearities and disturbances, a robustness-promoting regularizer refines the initial policy, while the adaptive section of the DeePO framework is enhanced with a forgetting factor to improve adaptation to time-varying dynamics.
The proposed DeePO+FL approach is evaluated through simulations and real-world experiments on an instrumented autonomous bicycle. Results demonstrate its superiority over the FL-only approach, achieving more precise tracking of the reference lean angle and lean rate.
\end{abstract}


\section{Introduction}
\IEEEPARstart{A}{n} autonomous bicycle is a bicycle, equipped with electric motors, sensors, algorithms, and control systems that allow the bicycle to navigate and operate without human intervention. Autonomous bicycles are an exciting area of research and development with numerous potential applications that can improve transportation, safety, and efficiency. In bicycle-sharing systems, autonomous bicycles can enhance the user experience by autonomously traveling to a person who has requested one, eliminating the need for individuals to walk toward the bicycle~\cite{sanchez2020autonomous}. Additionally, autonomous bicycles can streamline fleet management by enabling bicycles to autonomously navigate to charging stations for recharging. This eliminates the need for operators to manually collect, load, and transport bicycles to charging stations, making the process more efficient. Similar applications of the self-balancing feature of autonomous bicycles include steering assistants for individuals with limited physical capabilities \cite{alizadehsaravi2023bicycle}, among others.

Another notable application of autonomous bicycles is their ability to replace conventional bicycles in test tracks for evaluating the performance of various autonomous safety features in vehicles. Bicycles are often forced to share road segments with other motorized vehicles, which places cyclists at a higher risk of injury~\cite {kutela2021mining}. One way to reduce the risk is to use autonomous emergency braking (AEB) and autonomous emergency steering (AES) systems in motorized vehicles. The sensors in the vehicles detect and classify vulnerable road users (VRUs), including pedestrians and cyclists, and brakes or steers to avoid a collision. When organizations like EuroNCAP evaluate the AEB and AES systems on test tracks, a bicycle target placed on a moving platform is utilized~\footnote{\href{https://www.euroncap.com/en/vehicle-safety/the-ratings-explained/vulnerable-road-user-vru-protection/aeb-cyclist/}{https://www.euroncap.com/en/vehicle-safety/the-ratings-explained/vulnerable-road-user-vru-protection/aeb-cyclist/}}. Since the target is mounted on the platform, its movements are also constrained by its linear motion. An autonomous bicycle, which can better represent a cyclist's maneuvers and sometimes unpredictable behavior, would enhance the testing area, thereby improving the reliability of safety tests in vehicles. 

A bicycle is a nonlinear system that becomes self-stabilized at sufficient forward speed but remains unstable at lower velocities~\cite{Meijaard2007}. A human rider can employ three different control actions to maintain bicycle stability: adjusting the forward velocity, steering, and shifting their center of gravity to control the lean angle. At fixed speeds, similar to human control of a bicycle, an autonomous bicycle can be balanced either by regulating the steering angle \cite{Yeh2024, PerssonECC, defoort2009sliding} or by directly controlling the lean angle \cite{zhang2014, wang2017tracking}. However, the latter approach requires mounting a flywheel or a moving mass on the bicycle, significantly altering its appearance, which may be undesirable for certain applications, i.e., at test tracks of safety features of vehicles, where it is important that the autonomous bicycle resembles an ordinary bicycle. In addition, as revealed by Kooijman~\etal~\cite{kooijman2009} and later confirmed by Moore~\etal~\cite{moore2011}, controlling the lean angle has only a minor effect on balance at typical riding velocities. Therefore, this paper focuses on balancing the bicycle through steering control. Like human control at relatively high speeds, the bicycle should be steered in the direction of the fall, i.e., if it leans to the right, it must be steered to the right. Moreover, balancing the bicycle through steering control offers a more energy-efficient alternative than flywheels or a moving mass~\cite{rodriguez2017improving}. 
 
A wide range of control approaches have been proposed for balancing autonomous bicycles based on the simple principle of steering in the direction of the fall. For example, PID and LQR controllers have been designed based on a linearized model around the upright equilibrium of the bicycle and are evaluated around this equilibrium~\cite{PerssonECC}. Both PID and LQR offer simple design methodologies; however, as a natural consequence of linearization, their performance may degrade as the system deviates from the equilibrium point used for linearization. One way to address this drawback is using a nonlinear controller, such as the sliding mode controller (SMC). The second order SMC proposed in the work of Defoort~\etal~\cite{defoort2009sliding} is designed using a nonlinear point-mass model of a bicycle and evaluated through both simulations and on an instrumented bicycle riding on a bicycle roller. However, SMC may induce chattering in the control signal due to the switching nature of the control law, potentially leading to high actuation efforts from the motors~\cite{slotine1991applied}. 

Active disturbance rejection control (ADRC) is proposed in~\cite{Baquero-Suarez2018} to handle unmodeled noise and disturbances. This control design is based on the so-called Whipple model~\cite{Meijaard2007}, a fourth-order model linearized around small lean and steering angles. The model requires 25 physical parameters to be measured or estimated from the bicycle~\cite{kooijman2008}. The ADRC is first evaluated in simulation using a detailed CAD model of their bicycle, imported into ADAMS, and controlled through co-simulation with MATLAB. Next, experiments were conducted. The results demonstrate that the bicycle can balance in both simulations and on straight asphalt tracks under varying forward velocities. However, noticeable oscillation was observed in both the lean and steering angles. Robustness against speed variations and disturbances is also considered in the recent work of Yeh~\etal~\cite{Yeh2024}, in which a linear-parameter-varying (LPV) controller is designed based on a point mass model of a bicycle~\cite{Astrom2005}. The controller was evaluated in both simulations and experiments conducted on an instrumented bicycle. 

However, all these control methods rely heavily on an accurate system model. While in practice, the mathematical models of autonomous bicycles are reasonably well understood~\cite{Meijaard2007, bruni2020state},  the uncertainties, such as external disturbances (e.g., wind, street slope), and internal variations (e.g., changes in friction, mechanical and electrical couplings, and shifts in the center of gravity), can significantly degrade the performance of model-based approaches over time.

There are also direct approaches to bicycle control design that bypass the requirement of an explicit model. A notable instance is policy optimization (PO), an essential approach of modern reinforcement learning (RL)~\cite{choi2019toward, chung2017, weyrer2024path}. By computing the policy gradient from system trajectories, the PO updates the control policy with gradient descent methods. The work by Choi \etal~\cite{choi2019toward} employs Deep Deterministic Policy Gradient (DDPG) to learn a policy for controlling the bicycle's speed, lean angle, and steering torque. This approach is evaluated on a nonlinear bicycle model; however, its real-world transferability is not tested in \cite{choi2019toward}. The work by Tuyen and Chung~\cite{chung2017} evaluates DDPG on a miniature bicycle. However, the limited computational power of the hardware is insufficient for the control algorithm. Another deep RL algorithm is proposed for stabilizing a Whipple model of a bicycle and tracking a predefined path~\cite{weyrer2024path}. While the bicycle is balanced at varying velocities, they require time-consuming training and lack experimental validation. 

Recently, there has been a growing trend of direct data-driven control methods motivated by behavioral system theory and subspace methods~\cite{willems2005note,de2019formulas, Coulson2019, berberich2020data,FlorianReg2022, van2020data}. For example, the seminal work~\cite{de2019formulas} proposes a data-based parameterization for linear systems and reformulates the LQR problem as a convex program with a batch of persistently exciting data. Thus, the optimal LQR gain can be found without any explicit model or identification.  To enhance adaptability, our previous work~\cite{zhao2023data, zhao2024data} proposes a covariance parameterization for the LQR problem, based on which a data-enabled policy optimization (DeePO) method is developed to learn the LQR gain directly from online closed-loop data. In contrast to the previous policy optimization methods \cite{choi2019toward, chung2017, weyrer2024path} that require multiple long trajectories to compute a single policy gradient, the DeePO method uses the covariance of online data to update the policy sample-by-sample. Besides its sample efficiency, the DeePO method is also computationally efficient, performing only a single step of gradient descent to update the policy per time step. Under persistently exciting input, the DeePO algorithm is shown to have non-asymptotic convergence guarantees to the optimal LQR gain. The covariance parameterization or the DeePO method has been extended to linear parameter-varying control~\cite{mejari2024direct}, model-reference control~\cite{mejari2024bias}, output-feedback control, and validated in a power converter system via simulations~\cite{zhao2024direct}.

This paper proposes a unified DeePO framework for nonlinear bicycle control. To deal with the nonlinearities, we first design an output feedback linearization (FL) controller using well-established parametric bicycle models. This controller partially cancels the nonlinearities of the bicycle system while simultaneously stabilizing it. It is important to note that although the current state-of-the-art bicycle models are accurate, the model-based controllers may fail to achieve the desired performance due to unmodeled dynamics, parametric uncertainty, various disturbances, and time-varying dynamics (e.g., wind, varying friction in different environments, mechanical wear and tear, and coupling aging). To compensate for these disturbances and uncertainties, we integrate a DeePO controller on top of the FL layer. This allows us to adaptively fine-tune the feedback gains and effectively handle time-varying dynamics, disturbances, and unmodeled effects. Our approach is evaluated on both a realistic multi-body dynamic model of the bicycle and in hardware experiments conducted on an instrumented bicycle in an indoor environment. 

To the best of our knowledge, this study is the first to report a real-world implementation of DeePO. While previous works have explored DeePO in simulations~\cite{zhao2024data, zhao2024direct}, our study demonstrates its feasibility in a practical setting, validating its effectiveness in handling real-world nonlinearities, uncertainties, disturbances, and unmodeled dynamics. This contribution marks an important step toward bridging the gap between theory and real-world deployment of adaptive data-driven control methods.

\subsection{Statement of Contributions}
The contributions of this paper are summarized below.
\begin{itemize}  
    \item We introduce a unified control framework, combining FL with DeePO, for controlling a nonlinear autonomous bicycle.  
    \item We integrate a forgetting factor into the conventional DeePO framework to effectively handle time-varying dynamics.  
    \item This study is the first to apply DeePO in a real-world case study, specifically showcasing its effectiveness in controlling a bicycle through steering.
    \item Through both high-fidelity simulations and hardware experiments, we demonstrate that updating the feedback gains at every iteration of the DeePO algorithm's execution is not necessary. In fact, updating the feedback gain in DeePO with lower frequencies can still maintain or even improve control performance.
\end{itemize}  

\subsection{Organization}
The rest of the paper is organized as follows: Section~\ref{sec:2} formally states the control problem and provides an overview of our unified control framework. Section~\ref{sec:3} outlines the details of the DeePO algorithm, including a regularizer for the initial policy and incorporating a forgetting factor into the framework. Section~\ref{sec:4} presents the simulation and experimental results. Finally, Section~\ref{sec:5} presents the concluding remarks.

\subsection{Notation}
We use $I_n$ to denote the $n$-by-$n$ identity matrix. We use $\rho(\cdot)$ to denote the spectral radius of a square matrix. 
$A^\top$, $\text{Tr}(A)$, and $A^\dagger$ represent the transpose, trace, and pseudoinverse of matrix $A$, respectively. We use diag$(a, b, \dots, c)$ to denote a diagonal matrix with diagonal elements being $a,b,\dots,c$. The 2-norm of matrix $A$ is denoted $\|A\|$. We denote the continuous-time signal $x$ with $x(t)$ and discrete-time with $x\dt{t}$. 
\section{Autonomous bicycle control}
\label{sec:2}
This paper addresses the problem of designing a unified control method for balancing an autonomous bicycle while tracking a reference lean angle. First, we use a simple point-mass model to design an inner-loop FL control. Next, DeePO enhances system performance by adapting feedback gains based on input-output data from sensors mounted on the bicycle. 

\subsection{Bicycle Dynamics and Mathematical Modeling}
\begin{figure}[t]
    \centering
    \includegraphics[width=0.95\columnwidth]{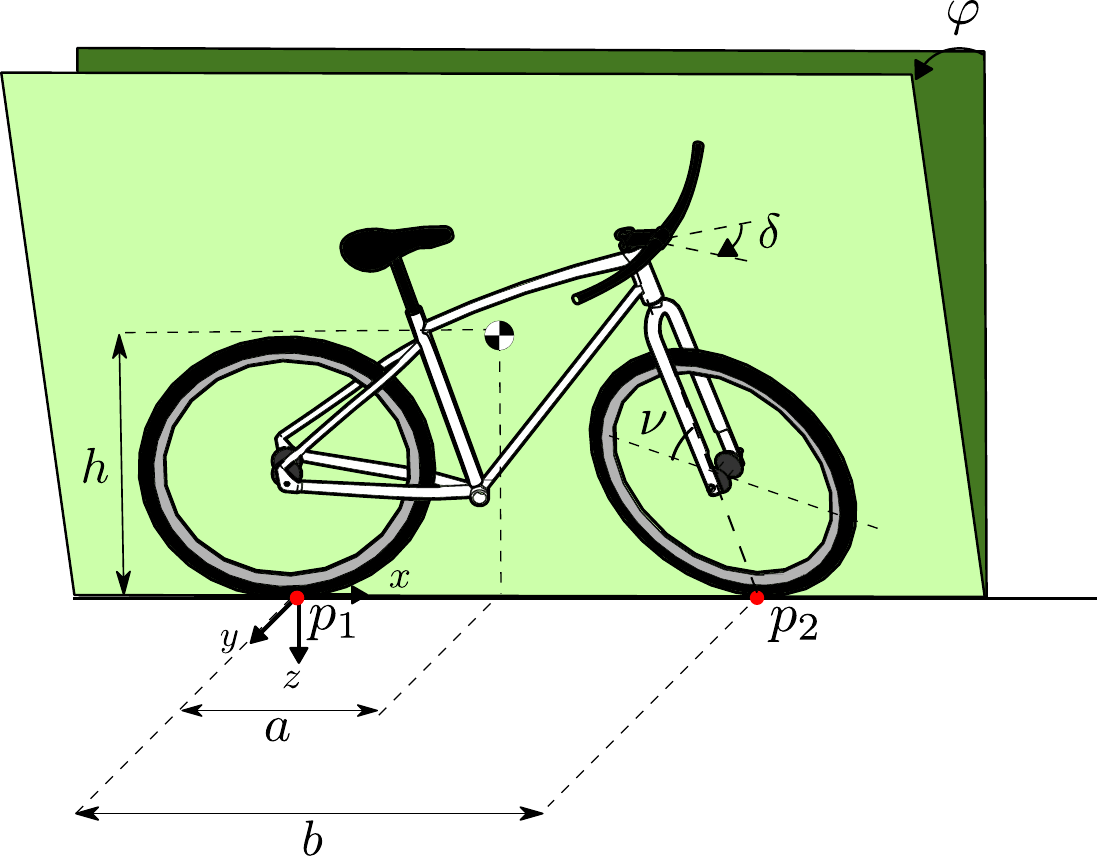}
    \caption{Illustration of the parameters used in the bicycle model in \eqref{eq:simpleModel}.}
    \label{fig:BikeModel}
\end{figure}

We consider a simple nonlinear model to represent the bicycle dynamics as in~\cite{Persson_2024}. 
\begin{equation}
    \label{eq:simpleModel}
    \begin{aligned}
        \ddot{\varphi}(t)&=\frac{g}{h}\sin\big(\varphi(t)\big)+ \frac{a}{bh}\cos\big(\varphi(t)\big)v\dot\delta(t) -\\
        & \left(\frac{1}{bh}- \frac{1}{b^2}\tan\big(\delta(t)\big) \tan\big(\varphi(t)\big)\right)\tan\big(\delta(t)\big)v^2,
    \end{aligned}
\end{equation}
where $\varphi(t)$, $\dot{\varphi}(t)$, $\delta(t)$, and $\dot{\delta}(t)$ represent the lean angle, lean rate, steering angle, and the controlled steering rate, respectively. The contact point between the rear wheel and the ground is denoted by $p_1$. Additionally, the vertical and horizontal distances between the bicycle's center of gravity and $p_1$ are denoted by $a$ and $h$, respectively. The wheelbase is denoted by $b$, while $g$ represents the gravitational constant, and $v$ represents the forward velocity. This model assumes a vertical steering axis, i.e., $\nu = \frac{\pi}{2}$, which results in zero trail. Furthermore, it is assumed that the steering axis can be controlled without delay and that the bicycle travels at a constant forward velocity. The visual representation of the parameters in \eqref{eq:simpleModel} is shown in Fig.~\ref{fig:BikeModel}.

\begin{figure}[t]
    \centering
    \includegraphics{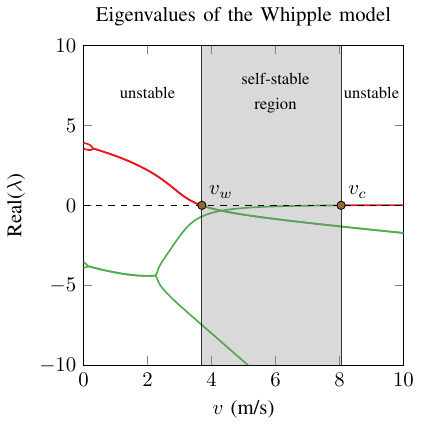}
    \caption{Stable and unstable regions of the instrumented bicycle, with the weave speed and capsize speed denoted by $v_w$ and $v_c$, respectively.}
    \label{fig:velAnalysis}
\end{figure}

A bicycle is self-stabilized between the so-called \textit{weave speed} and \textit{capsize speed}. By analyzing the eigenvalues of the Whipple model and identifying the region where they are all negative, the self-stable region of a bicycle can be localized~\cite{kooijman2008}. A similar eigenvalue analysis for the instrumented bicycle considered in this paper was previously conducted, where the $25$ parameters required for the Whipple model were measured~\cite{persson2023control}. Based on this analysis, we focus on forward speeds of approximately $8$ km/h ($2.22$ m/s), below the weave speed, as shown in Fig.~\ref{fig:velAnalysis}. Thus, the system we aim to control is open-loop unstable, nonlinear, and non-holonomic, presenting a challenging control problem. Due to the system's inherent instability, applying a persistently exciting input without additional stabilization can lead to a loss of balance and cause the system to diverge. Specifically, an uncontrolled persistently exciting input could destabilize steering actions, such as turning left while leaning right, making it impractical to rely solely on such input for collecting persistently exciting data. In the following, we present an FL controller that balances the bicycle, simplifying the acquisition of persistently exciting data and mitigating some of the system's nonlinearities.

\subsection{Control overview}
A common approach for realizing a persistently exciting input is to utilize a random signal~\cite{alsalti2023design}. However, this approach is not directly applicable as it jeopardizes the bicycle's balance. Instead, we pre-stabilize the bicycle with an inner control loop using FL. Since the relative degree between the output and the number of states in the model given by \eqref{eq:simpleModel} does not match, the system can only be partially linearized using output FL~\cite{slotine1991applied}. 

If we choose $x = \begin{bmatrix}
    x_1, & x_2, & x_3
\end{bmatrix} = \begin{bmatrix}
    \varphi(t), & \dot{\varphi}(t), & \delta(t)
\end{bmatrix}, $ $y(t) = \varphi(t)$, $\dot{y}(t) = \dot{\varphi}(t)$, and represent the reference output as $y_r = [y_r(t), \dot{y}_r(t), \ddot{y}_r(t)]$, we can express the considered FL control law as:

\begin{equation}
    u(t) = \dot{\delta}(t) = \frac{1}{p(x)}(w - f(x)),
    \label{eq:FLinput}
\end{equation}

where 
\begin{align}
\label{eq:control}
    f(x)&= - \left(\frac{1}{bh}- \frac{1}{b^2}\tan\big(x_3\big) \tan\big(x_1\big)\right)\tan\big(x_3\big)v^2 \nonumber \\
      &\phantom{=} + \frac{g}{h}\sin\big(x_1\big) \nonumber \\
    p(x) & =\frac{a}{bh}\cos\big(x_1\big)v, \nonumber \\
    w & =  \ddot y_r(t) + k_1\left(\dot y_r(t)-\dot y(t)\right) +k_2 \left(y_r(t)-y(t)\right),\\ \nonumber
\end{align}
with appropriate choices of $k_1 > 0$ and $k_2 > 0$ to partially compensate for the system's nonlinearities. However, the steering angle $\delta(t)$ remains an internal state that is not directly linearized, meaning some nonlinear dynamics persist. In particular, terms involving $\tan(\delta(t))$ introduce coupling effects that remain even after feedback linearization. Additionally, since the steering angle evolves according to $u = \dot{\delta}$, it can drift over time uncontrolled, requiring further regulation to prevent undesired effects on system stability.
Furthermore, the proposed FL controller is designed based on continuous-time dynamics, with the model and control parameters provided in \eqref{eq:simpleModel} and Table~\ref{tab:modParam}, respectively. However, in practice, we implement it using a sampled-data approach with a hold mechanism, which may introduce inaccuracies and lead to performance degradation due to the discrete nature of the implementation. This discrete approach may not fully capture the continuous dynamics of the system~\cite{kimber1991sampled, grizzle1988feedback}. Moreover, parameters in~\eqref{eq:FLinput} and~\eqref{eq:control} are subject to parametric uncertainty, resulting in an inaccurate canceling of nonlinearties.  Nevertheless, we demonstrate that the potential limitations of the FL controller can be mitigated by incorporating DeePO.

\begin{table}[b]
\centering
\caption{Model and control parameters for FL control}
\label{tab:modParam}
\begin{tabular}{@{}llll@{}}
\toprule
\textbf{Parameter}  & \textbf{Symbol} & \textbf{Value} & \textbf{Unit} \\ \midrule
CoG w.r.t $p_1$ (x) & $a$             & $0.550$        & m             \\
CoG w.r.t $p_1$ (z) & $h$             & $0.700$        & m             \\
Wheelbase           & $b$             & $1.200$        & m             \\
Gravity             & $g$             & $9.82$         & m/s$^2$       \\
$k_1$               & -               & 1              & -             \\
$k_2$               & -               & 6              & -             \\ \bottomrule
\end{tabular}
\end{table}

\begin{figure}[t]
    \centering
    \includegraphics{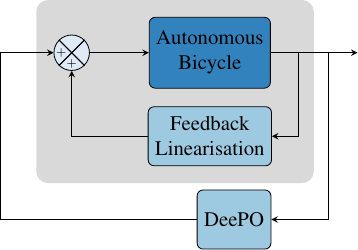}
    \caption{Control overview}
    \label{fig:controlOverview}
\end{figure}

The proposed FL controller functions as an inner control loop to stabilize the bicycle, enabling the use of an additive random signal as either a persistently exciting input or a performance enhancing adaptive control. In the remainder of the paper, we consider the autonomous bicycle with FL as our target system to control by DeePO, as highlighted by the gray box in Fig.~\ref{fig:controlOverview}. With this stable inner-loop system in place, we shift our focus to enhancing performance and compensating for the remaining nonlinearities using an adaptive, direct data-driven control approach in the outer loop.

\section{Data-enabled policy optimization for autonomous bicycle  control}
\label{sec:3}
This section first describes a brief overview of the linear quadratic regulator (LQR). Then, we propose a data-enabled policy optimization with a forgetting factor for adaptive learning of the LQR based on~\cite{zhao2023data, zhao2024data}. 

\subsection{The linear quadratic regulator}
	Consider a linear time-invariant system
	\begin{equation}\label{equ:sys}
	\left\{\begin{aligned}
	x\dt{t+1} & =A x\dt{t}+B u\dt{t}+w\dt{t} \\
	h\dt{t} & =\begin{bmatrix}
	Q^{1 / 2} & 0 \\
	0 & R^{1 / 2}
	\end{bmatrix}
	\begin{bmatrix}
	x\dt{t}  \\
	u\dt{t} 
	\end{bmatrix}
	\end{aligned}\right. .
	\end{equation}
	Here, $x\dt{t}$ is the state, $u\dt{t}$ is the control input, $w\dt{t}$ represents the noise, and $h\dt{t}$ is the performance signal of interest, $(A,B)$ are controllable, and the weighting matrices $(Q, R)$ are positive definite. 
	
    The LQR problem aims to find an optimal state-feedback gain $K\in \mathbb{R}^{m\times n}$ that minimizes the $\mathcal{H}_2$-norm of the transfer function $\mathscr{T}(K):w \rightarrow h$ of the closed-loop system
	\begin{equation}
	\begin{bmatrix}
	x\dt{t+1}  \\
	h\dt{t} 
	\end{bmatrix}=\begin{bmatrix}
	A+BK & I_n \\
	\hline \begin{bmatrix}
	Q^{1 / 2} \\
	R^{1 / 2} K
	\end{bmatrix} & 0
	\end{bmatrix}\begin{bmatrix}
	x\dt{t}  \\
	w\dt{t} 
	\end{bmatrix}.
	\end{equation}
	When $A+BK$ is stable, it holds that \cite{anderson2007optimal}
	\begin{equation}\label{equ:transfer}
	\|\mathscr{T}(K)\|_2^2  = \text{Tr}((Q+K^{\top}RK)\Sigma_K)=:C(K),
	\end{equation}
	where $\Sigma_K$ is the closed-loop state covariance matrix obtained as the positive definite solution to the Lyapunov equation
	\begin{equation}\label{equ:Sigma}
	\Sigma_K = I_n + (A+BK)\Sigma_K (A+BK)^{\top}.
	\end{equation}
	We refer to $C(K)$ as the LQR cost and to (\ref{equ:transfer})-(\ref{equ:Sigma}) as a \textit{policy parameterization} of the LQR.

The optimal LQR gain $K^*$ is unique and can be found by, e.g., solving an algebraic Riccati equation with $(A,B)$~\cite{anderson2007optimal}. When $(A,B)$ is unknown, data-driven methods can be used to learn the LQR gain from input-state data. In the sequel, we propose a covariance parameterization method for direct data-driven learning of the LQR.

\subsection{Data-driven covariance parametrization of the LQR with exponential weighted data}\label{sec:3-B}
Consider the $t$-long time series of states, inputs, noises, and successor states
\begin{align}
X_{0,t} &:= \begin{bmatrix}
x\dt{0} & x\dt{1} & \dots& x\dt{t-1} 
\end{bmatrix}\in \mathbb{R}^{n\times t},\nonumber\\
U_{0,t} &:= \begin{bmatrix}
u\dt{0} & u\dt{1} & \dots& u\dt{t-1} 
\end{bmatrix}\in \mathbb{R}^{m\times t}, \label{equ:dataMat}\\
W_{0,t} &:= \begin{bmatrix}
w\dt{0} & w\dt{1} & \dots& w\dt{t-1} 
\end{bmatrix}\in \mathbb{R}^{n\times t}, \nonumber\\
X_{1,t} &:= \begin{bmatrix}
x\dt{1} & x\dt{2} & \dots& x\dt{t} 
\end{bmatrix}\in \mathbb{R}^{n\times t}, \nonumber
\end{align}
which satisfy the system dynamics
\begin{equation}\label{equ:dynamics}
X_{1,t} = AX_{0,t}+ BU_{0,t} + W_{0,t}.
\end{equation}

Assume that the data is {\em persistently exciting (PE)} \cite{willems2005note}, i.e., the block matrix of input and state data
\begin{equation}
D_{0,t} := 
\begin{bmatrix}
    U_{0,t} \\
    X_{0,t}
\end{bmatrix}
\end{equation}
has full row rank
\begin{equation}\label{equ:rank}
\text{rank}(D_{0,t}) = m+n.
\end{equation}
Define the covariance of exponentially weighted data as
\begin{equation}
    \Phi_{t} := \frac{1}{t}D_{0,t} S_{\lambda} D_{0,t}^{\top},
\end{equation}
where $\lambda \in (0,1)$ is a forgetting factor and $S_{\lambda} := \text{diag}\{ \lambda^{t-1}, \lambda^{t-2},\dots, 1\}$. Compared with \cite{zhao2024data}, the forgetting factor here makes the weight of past data decay exponentially, such that the sample covariance can also reflect and adapt to the behavior of time-varying or nonlinear systems.

Since $D_{0,t}$ has full row rank and $S_{\lambda}\succ 0$, the covariance matrix is positive definite, i.e., $\Phi_{t} \succ 0$. Then, for any gain $K$, there exist a matrix $V$ such that
\begin{equation}\label{equ:forget}
\begin{bmatrix}
K \\
I_n
\end{bmatrix}=  \Phi_{t} V.
\end{equation}
We refer to \eqref{equ:forget} as the \textit{covariance parameterization} with exponentially weighted data and to $V\in \mathbb{R}^{(n+m)\times n}$ as the \textit{parameterized policy}.

With \eqref{equ:forget}, the LQR problem \eqref{equ:transfer}-\eqref{equ:Sigma} can be expressed by raw data matrices $(X_{0,t}, U_{0,t}, X_{1,t})$ and the optimization matrix $V$. For brevity, let $\overline{X}_{0,t}= X_{0,t}S_{\lambda}D_{0,t}^{\top}/t$ and $\overline{U}_{0,t}=  U_{0,t}S_{\lambda}D_{0,t}^{\top}/t$ be a partition of $\Phi_t$, and let
$\overline{W}_{0,t}=  W_{0,t}S_{\lambda}D_{0,t}^{\top}/t$ be the noise-state-input covariance, and finally define the covariance with respect to the successor state as $\overline{X}_{1,t}=  X_{1,t}S_{\lambda}D_{0,t}^{\top}/t$.
Then, the closed-loop matrix can be written as
\begin{equation}
A+BK=[B,A]\begin{bmatrix}
K \\
I_n
\end{bmatrix}\overset{\eqref{equ:forget}}{=}[B,A]\Phi_t V\overset{\eqref{equ:dynamics}}{=}(\overline{X}_{1,t} - \overline{W}_{0,t})V.
\end{equation}
Following the certainty-equivalence principle~\cite{dorfler2021certainty}, we disregard the
unmeasurable $\overline{W}_{0,t}$ for the design and use $\overline{X}_{1,t}V$ as the closed-loop matrix. After substituting $A+BK$  with $\overline{X}_{1,t}V$ in (\ref{equ:transfer})-(\ref{equ:Sigma}) and leveraging \eqref{equ:forget}, the LQR problem becomes 
\begin{equation}\label{prob:equiV}
\begin{aligned}
&\mathop{\text {minimize}}\limits_{V}~J_t(V) :=\text{Tr}\left((Q+V^{\top}\overline{U}_{0,t}^{\top}R\overline{U}_{0,t}V)\Sigma_t(V)\right),\\
&\text{subject to}~ ~\overline{X}_{0,t}V= I_n,
\end{aligned}
\end{equation}
where $\Sigma_t(V) = I_n + \overline{X}_{1,t}V\Sigma_t(V) V^{\top}\overline{X}_{1,t}^{\top}$ is a covariance parameterization of \eqref{equ:Sigma},
and the original gain matrix can be recovered as $K = \overline{U}_{0,t}V$. We refer to (\ref{prob:equiV}) as the covariance-parameterized LQR problem, which is direct data-driven and does not involve any explicit SysID.

\subsection{Data-enabled policy optimization for adaptive LQR control with exponentially weighted data}
In previous work \cite{zhao2023data,zhao2024data}, a data-enabled policy optimization (DeePO) method for direct adaptive learning of the LQR was proposed, where the control policy is parameterized by sample covariance and updated recursively using gradient methods. In this subsection, we propose a DeePO algorithm based on our covariance parameterization with exponentially weighted data \eqref{equ:forget}, detailed in Algorithm \ref{alg:deepo}.

Algorithm \ref{alg:deepo} alternates between control (line 2) and policy update (lines 3-6).
The DeePO algorithm uses online gradient descent of \eqref{prob:equiV} to recursively update $V$. 
At time $t$, we apply the linear state feedback policy $u\dt{t}=K_tx\dt{t}  + e\dt{t} $ for control and observe the new state $x\dt{t+1} $, where $e\dt{t}$ is a probing noise used to ensure the PE rank condition \eqref{equ:rank}. To update the policy, we first use $(X_{0,t+1}, U_{0,t+1}, X_{1,t+1})$ to formulate the covariance-parameterized LQR problem \eqref{prob:equiV}. Then, instead of solving this optimization problem optimality, we only take a single step of projected gradient descent towards its solution in \eqref{equ:pro_gd}. Here, the projection  
\begin{equation}
\Pi_{\overline{X}_{0,t+1}}: = I_{n+m}-\overline{X}_{0,t+1}^{\dagger}\overline{X}_{0,t+1}
\end{equation}
onto the nullspace of $\overline{X}_{0,t+1}$ is to ensure the subspace constraint in \eqref{prob:equiV}.
 Define the feasible set of \eqref{prob:equiV} (i.e., the set of stable closed-loop matrices) as $\mathcal{S}_t:= \{V\mid \overline{X}_{0,t}V =I_n,  \rho (\overline{X}_{1,t}V)<1\}$. Then, the gradient can be computed as follows.
\begin{lemma}[\cite{zhao2024data}]\label{lem:gradient}
For $V\in \mathcal{S}_t$, the gradient of $J_t(V)$ with respect to $V$ is given by
	\begin{equation}\label{equ:pg}
	\nabla J_t(V) = 2 \left(\overline{U}_{0,t}^{\top}R\overline{U}_{0,t}+\overline{X}_{1,t}^{\top}P_t\overline{X}_{1,t}\right)V \Sigma_t(V),
	\end{equation}
	where $P_t$ satisfies the Lyapunov equation 
	\begin{equation}
	P_t = Q + V^{\top}\overline{U}_{0,t}^{\top}R\overline{U}_{0,t}V + V^{\top}\overline{X}_{1,t}^{\top}P_t\overline{X}_{1,t}V.
	\end{equation}
\end{lemma}

Algorithm \ref{alg:deepo} is \textit{direct and adaptive} in the sense that it directly uses online closed-loop data to update the policy. Thanks to the forgetting factor, it can rapidly adapt to changes in system behavior reflected in the data. 
As in \cite{zhao2024data}, Algorithm \ref{alg:deepo} can also be implemented recursively. We write the sample covariance  recursively as
\begin{equation}\label{equ:recur}
    \Phi_{t+1} = \frac{\lambda t}{t+1} \Phi_{t} + \frac{1}{t+1} \phi_t\phi_t^{\top},
\end{equation}
where $\phi_t = [u_t^\top, x_t^\top]^\top$. By the Sherman-Morrison formula~\cite{sherman1950adjustment}, its inverse $\Phi_{t+1}^{-1}$ satisfies
\begin{equation}
    \Phi_{t+1}^{-1} = \frac{t+1}{\lambda t}\left(\Phi_{t}^{-1} - \frac{\Phi_{t}^{-1}\phi_t\phi_t^{\top}\Phi_{t}^{-1}}{\lambda t+\phi_t^{\top}\Phi_{t}^{-1}\phi_t}\right).
\end{equation}
Furthermore, the rank-one update of the parameterized policy is given by
\begin{align}
V_{t+1}&= \frac{t+1}{t}\left(\Phi_{t}^{-1} - \frac{\Phi_{t}^{-1}\phi_t\phi_t^{\top}\Phi_{t}^{-1}}{t+\phi_t^{\top}\Phi_{t}^{-1}\phi_t}\right) \Phi_{t} V_{t}' \nonumber \\ 
&= \frac{t+1}{\lambda t} \left(V_{t}' - \frac{\Phi_{t}^{-1}\phi_t\phi_t^{\top}V_{t}'}{\lambda t+\phi_t^{\top}\Phi_{t}^{-1}\phi_t}\right),
\end{align}
where $\Phi_{t}^{-1}$ and $V_{t}'$ are given from the last iteration.

\begin{remark}
Using the forgetting factor $\lambda$ may asymptotically lead to failure of the rank-one update as time tends to infinity. To see this, we notice that the covariance update in
\eqref{equ:recur} satisfies stable linear dynamics. This implies a loss of persistency of excitation and $\Phi_t$ will tend to zero, and hence $\Phi^{-1}_t$ will grow to infinity. A simple remedy is to reset the covariance $\Lambda_{t}$  occasionally, i.e., set $\Lambda_{t}=I_{n+m}, \forall t\in \{T,2T,\dots\}$. Since the autonomous bicycle operates only for a finite time, we do not reset the covariance in our subsequent experiments in Section \ref{sec:4}. Another approach is to use sliding window data rather than exponentially weighted data for the covariance parameterization \eqref{equ:forget}, where a key is to select an optimal window size to balance data informativity and adaptation efficiency. We leave this exploration to future work.
\qed
\end{remark}

\begin{remark}
    The stepsize $\eta_t$ should be set according to the signal-to-noise ratio (SNR) of online data. For example, when the SNR is large, we are confident with the gradient direction, and the stepsize can be chosen more aggressively; on the contrary, when the SNR is small, the stepsize should be small to prevent the policy from moving out of the stability region. To this end, we set the stepsize as 
    \begin{equation}    
    \eta_t = \frac{\eta_0}{\left\|\overline{U}_{0,t}\Pi_{\overline{X}_{0,t}}\overline{U}_{0,t}^{\top}\right\|}, ~t\geq t_0,
    \end{equation}
    where $\eta_0$ is a constant, and the denominator is used to quantify the SNR. Another motivation of the denominator is from \cite[Lemma 3]{kang2024linear}, which reveals the equivalence between data-enabled and model-based policy gradients up to the data matrix $\overline{U}_{0,t}\Pi_{\overline{X}_{0,t}}\overline{U}_{0,t}^{\top}$. \qed
\end{remark}

\begin{algorithm}[t]
	\caption{DeePO for direct adaptive LQR control}
	\label{alg:deepo}
	\begin{algorithmic}[1]
		\Require Offline data $(X_{0,t_0}, U_{0,t_0}, X_{1,t_0})$, an initial policy $K_{t_0}$, and a stepsize $\eta$.
		\For{$t=t_0,t_0+1,\dots$}
		\State Apply $u\dt{t} =K_tx\dt{t}  + e\dt{t} $ and observe $x\dt{t+1} $.
        \State Update covariance matrices $\Phi_{t+1}$ and $\overline{X}_{1,t+1}$.
		\State \textbf{Policy parameterization:} given $K_{t}$, solve $V_{t+1}$ via 
		$$
		V_{t+1} =\Phi_{t+1}^{-1} \begin{bmatrix}
		K_{t} \\
		I_n
		\end{bmatrix}.
		$$
		\State \textbf{Update of the parameterized policy:} perform one-step projected gradient descent
		\begin{equation}\label{equ:pro_gd}
		V_{t+1}' = V_{t+1} - \eta_t  \Pi_{\overline{X}_{0,t+1}} \nabla J_{t+1}(V_{t+1}),
		\end{equation} 
		where the gradient $\nabla J_{t+1}(V_{t+1})$ is given by Lemma \ref{lem:gradient}.
		\State \textbf{Gain update:} update the control gain by 
		$$
		K_{t+1} = \overline{U}_{0,t+1}V_{t+1}'.
		$$
		\EndFor	
	\end{algorithmic}
\end{algorithm}

Algorithm \ref{alg:deepo} requires the initial policy to be stabilizing. A potential approach is to solve the  covariance-parameterized LQR \eqref{prob:equiV} with the offline data $(X_{0,t_0}, U_{0,t_0}, X_{1,t_0})$. However, due to the nonlinearity in the system dynamics of the autonomous bicycle, the solution of covariance-parameterized LQR may be destabilizing. Next, we propose a robustness-promoting regularizer for the covariance parameterization to obtain a stabilizing initial policy.
 
\subsection{Learning an initial stabilizing policy using robustness promoting regularization}

	The feasibility of the covariance-parameterized LQR problem \eqref{prob:equiV} depends on that of the Lyapunov equation
	\begin{equation}\label{equ:lyap}
		\Sigma = I_n + \overline{X}_1V\Sigma V^{\top}\overline{X}_1^{\top},
	\end{equation}
	where $\overline{X}_1V$ is regarded as the closed-loop matrix. However, having assumed certainty-equivalence by the covariance parameterization \eqref{equ:forget} and the relation $A+BK = (\overline{X}_1 - \overline{W}_0)V$, the Lyapunov equation that should be met is
	\begin{equation}\label{equ:true_lyap}
		\Sigma = I_n + (\overline{X}_1 - \overline{W}_0)V\Sigma V^{\top}(\overline{X}_1 - \overline{W}_0)^{\top}.
	\end{equation}
	The gap between the right-hand side of \eqref{equ:lyap} and \eqref{equ:true_lyap} is
	\begin{equation}\label{equ:diff}
		\begin{aligned}
		&\overline{W}_0 V\Sigma V^{\top}  \overline{W}_0^{\top} - \overline{W}_0 V\Sigma V^{\top} \overline{X}_1^{\top} -   \overline{X}_1V\Sigma V^{\top} \overline{W}_0^{\top} \\
		&= \frac{1}{t^2}  W_0D_0^{\top}V\Sigma V^{\top}D_0W_0^{\top} \\
		&-\frac{1}{t^2}(  W_0D_0^{\top}V\Sigma V^{\top}D_0X_1^{\top} +   X_1D_0^{\top}V\Sigma V^{\top}D_0W_0^{\top}).
		\end{aligned}
	\end{equation}
	To reduce the gap, it suffices to make $\text{Tr}(D_0^{\top}V\Sigma V^{\top}D_0/t)$ small. To this end, we introduce the regularizer $\text{Tr}(V\Sigma V^{\top}\Phi)$ to the covariance-parameterized LQR problem \eqref{prob:equiV}, leading to
	\begin{equation}\label{prob:regu}
	\begin{aligned}
	&\mathop{\text {minimize}}\limits_{V, \Sigma\succeq 0}~ J_t(V) + \gamma\text{Tr}(V\Sigma V^{\top}\Phi),\\
	&\text{subject to}~ ~\Sigma = I_n + \overline{X}_1V\Sigma V^{\top}\overline{X}_1^{\top},\overline{X}_0V= I_n
	\end{aligned}
	\end{equation}
	with gain matrix $K = \overline{U}_0V$, where $\gamma>0$ is the regularization coefficient. We refer to \eqref{prob:regu} as the regularized covariance parameterization of the LQR problem.

    To obtain an initial stabilizing policy for Algorithm \ref{alg:deepo}, we solve \eqref{prob:regu} with offline data $(X_{0,t_0}, U_{0,t_0}, X_{1,t_0})$.

\subsection{Control gain update rate}
Rapid changes in an adaptive control policy, $K_t$ can potentially induce oscillations and, in the worst case, render the system unstable~\cite{landau2011adaptive}. Moreover, the control policy at certain time intervals may be significantly influenced by measurement noise, meaning that updates could be driven more by noise than by the actual system dynamics. 

To address these potential issues, we propose updating the DeePO control gain less frequently than the sampling frequency. To regulate the update frequency, we introduce the parameter $\xi$, which determines the intervals at which the controller in line 6 of Algorithm~\ref{alg:deepo} is updated. For instance, if $\xi = 1$, the control gain is updated at every iteration, whereas if $\xi = 100$, the gain is updated every $100$ iterations.

\section{Simulations and experiments}
\label{sec:4}
\definecolor{c1}{RGB}{27,158,119}
\definecolor{c2}{RGB}{217,95,2}
\definecolor{c5}{RGB}{117,112,179}
\definecolor{c4}{RGB}{231,41,138}
\definecolor{c3}{RGB}{102,166,30}
\definecolor{c6}{RGB}{230,171,2}

\colorlet{c3a}{c3!100}
\colorlet{c3b}{c3!80}
\colorlet{c3c}{c3!60}
\colorlet{c3d}{c3!40}

\colorlet{c4a}{c4!100}
\colorlet{c4b}{c4!80}
\colorlet{c4c}{c4!60}
\colorlet{c4d}{c4!40}

\colorlet{c32a}{c3!100}
\colorlet{c32b}{c3!88}
\colorlet{c32c}{c3!76}
\colorlet{c32d}{c3!64}
\colorlet{c32e}{c3!52}
\colorlet{c32f}{c3!40}

\colorlet{c42a}{c4!100}
\colorlet{c42b}{c4!88}
\colorlet{c42c}{c4!76}
\colorlet{c42d}{c4!64}
\colorlet{c42e}{c4!52}
\colorlet{c42f}{c4!40}

In this section, we first provide details of the instrumented bicycle used in the experiments. Next, we describe the simulation setup, followed by the results obtained from the simulations. Finally, we present the details of the experiments and the corresponding results.

\subsection{Instrumented Bicycle}
\begin{figure}[t]
    \centering
    \includegraphics{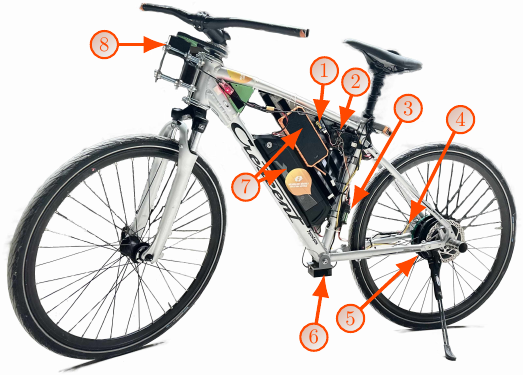}
    \begin{tabular}{@{}llll@{}}
        \toprule
        \multicolumn{4}{c}{\textbf{Hardware}} \\ \midrule
        \includegraphics[height=0.5cm]{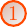}   & RC receiver    
        & \includegraphics[height=0.5cm]{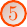}     & Bafang RM G040.250.DC    \\
        \includegraphics[height=0.5cm]{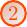}      & Raspberry Pi 4b     & 
        \includegraphics[height=0.5cm]{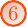} &   Xsens MTi-7  \\
        \includegraphics[height=0.5cm]{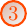}   & ESC    
        & \includegraphics[height=0.5cm]{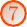}     & Batteries    \\
        \includegraphics[height=0.5cm]{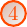}      & Hall sensor     & 
        \includegraphics[height=0.5cm]{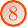}      & Dynamixel XH540-W270-T \\ \bottomrule
        \end{tabular}
    \caption{Instrumented bicycle used in the experiments.}
    \label{fig:expPlatform}
\end{figure}

The bicycle we consider in experiments is a men's model of an electric bicycle, as shown in Fig.~\ref{fig:expPlatform}. The factory-installed rear wheel motor and the battery mounted on the main frame are used to drive the bicycle forward. The rear wheel is controlled through a Phoenix Edge HV 60 AMP Electronic Speed Controller (ESC), and the rear wheel velocity is estimated using 12 evenly distributed magnets on the rear wheel and a Hall sensor. A PI controller is manually tuned to maintain an approximately constant forward velocity. An Xsens MTi-7 GNSS/INS at the bottom bracket shell measures the lean angle and rate.

Furthermore, a Dynamixel XH540-W270-T servo with a cog belt is mounted on the bicycle's main frame to control the handlebar. The servo is controlled through a velocity command, $u\dt{t}=\dot{\delta}\dt{t}$ rad/s, using an integer between $-167$ and $167$ with a resolution of approximately $0.024$ rad/s per unit. The control signal is saturated at $\pm4$ rad/s. The control algorithms are implemented, and the data is processed using ROS2 Humble running on a Raspberry Pi 4b with Ubuntu 20.04. Finally, a radio controller (RC) lets the operator send wireless commands to the RC receiver mounted on the bicycle. 

\subsection{Simulation setup}
\begin{figure}[t]
    \centering
    \includegraphics[width=0.48\textwidth]{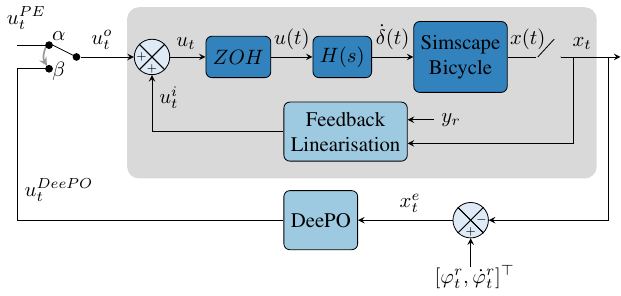}
    \caption{Control setup where the gray box represents the system controlled using DeePO.}
    \label{fig:ControlStructure}
\end{figure}

A CAD model of the instrumented bicycle was designed using SolidWorks. The CAD model is imported into MathWorks Simscape and controlled through Simulink. The rear and front wheels are connected to the mainframe through revolute joints, where the rear joint is actuated and given a constant speed corresponding to a forward velocity of $8$ km/h. A third revolute joint connects the steering axis to the bicycle's mainframe and is actuated through the control signal $u(t) = \dot\delta(t)$. The steering dynamics are modeled using an identified steering step response matching procedure~\cite{PerssonECC}, from the control signal $u(t)$ to the steering rate $\dot\delta(t)$ the resulting transfer function is: 
\begin{equation}
    H(s) = \frac{100+s}{100}.
\end{equation}
The transfer function is placed in series with the bicycle model, as shown in Fig.~\ref{fig:ControlStructure}. The control signal is saturated at $\pm 4$ rad/s. The figure also highlights that the input signal, $u\dt{t}$, is composed of two parts. First, an inner loop control signal $u\dti{t}{i}$ from the FL controller with the parameters given in Table \ref{tab:modParam}, and second, an outer control loop signal $u\dti{t}{o}$ which originates from a persistently excited input $u\dti{t}{PE} = \mathcal{N}(0,\sigma_{PE})$ with $\sigma_{PE}=0.2$ rad/s when the switch is in position $\alpha$, and from DeePO when the switch is in position $\beta$. The lean angle, lean rate, and steering angle measurements are induced with zero-mean, normally distributed noise using $\sigma_{\varphi} = \sigma_{\delta}=0.5$ deg and $\sigma_{\dot{\varphi}} =0.5$ deg/s. The forward velocity is set to a constant value of $8$ km/h.

The tracking error is used as the state vector, i.e., $x\dti{t}{e} = [\varphi\dti{t}{r} - \varphi\dt{t}, \dot\varphi\dti{t}{r} -\dot\varphi\dt{t}]^\top$. The state variables and the persistently exciting input signal are sampled at $100$ Hz. The same execution rate is used for DeePO and the FL controller, generating a new control signal for the steering motor every $0.01$ s.

An initial simulation is performed with the switch in Fig.~\ref{fig:ControlStructure} in position $\alpha$, and the bicycle tracks a lean angle reference $\varphi\dti{t}{r}=0$ and its derivative $\dot\varphi\dti{t}{r}=0$ for 10 seconds.  We collect $T=200$ samples of the acquired data to construct $U_{0,T}, X_{0,T}, X_{1,T}$ using \eqref{equ:dataMat}. An initial policy is obtained by solving the regularized covariance-parameterized LQR problem~\eqref{prob:regu} using CVX~\cite{cvx} with $Q=\diag{[1,0.01]}$ and $R=10^{-4}$. Higher values of $\gamma$ promote robustness against uncertainties in the system, while lower values prioritize performance. In our simulations, we set it $\gamma = 1$, which balances robustness and performance in our initial control policy. 
 
 Next, the switch in Fig.~\ref{fig:ControlStructure} is set to position $\beta$, and Algorithm~\ref{alg:deepo} is used to update the control policy at every time sample, i.e., $\xi=1$ and using a forgetting factor $\lambda = 1-10^{-4}$, and learning rate $\eta=10^{-3}$. To ensure a persistently exciting input, the probing noise $e\dt{t}$ is a zero-mean normally distributed random number, which is added to the DeePO output and constructs the input to the system as:
\begin{equation}
    u\dti{t}{DeePO} = u\dti{t}{DeePO} + e\dt{t},
\end{equation}
where $e\dt{t}=\mathcal{N}(0,0.2u\dti{t}{DeePO})$. Moreover, the bicycle tracks a time-varying reference, as shown in Fig.~\ref{fig:simResults}.

To evaluate the update rate for the gain update in Algorithm~\ref{alg:deepo} (line 6), multiple simulations are conducted where $\xi$ varies while the rest of the parameters are kept fixed. Four different update rates for the gain update are considered: $\xi=1$, $10$, $50$, and $100$. Moreover, one simulation is conducted with only the FL controller as a baseline. The forgetting factor is evaluated in a similar fashion using $\lambda = 1-10^{-\zeta}$ with $\zeta=\infty$, $2$, $3$, $4$, $5$, $6$, where $\zeta=\infty$ corresponds to a controller without a forgetting factor. In these simulations, the control update rate is set to $\xi=1$, and the rest of the parameters are kept at their initial values. The performance of the controllers is evaluated using the integrated squared error of the lean angle and the lean rate with respect to their respective reference values: 
\begin{equation}
    \text{ISE}_\varphi = \sum_{i=0}^t (\varphi\dt{i}-\varphi\dti{i}{r})^2,\quad \text{ISE}_{\dot{\varphi}} = \sum_{i=0}^t (\dot{\varphi}\dt{i}-\dot{\varphi}\dti{i}{r})^2.
\end{equation}

\subsection{Simulation results}
\begin{figure}[t]
    \centering
    \includegraphics{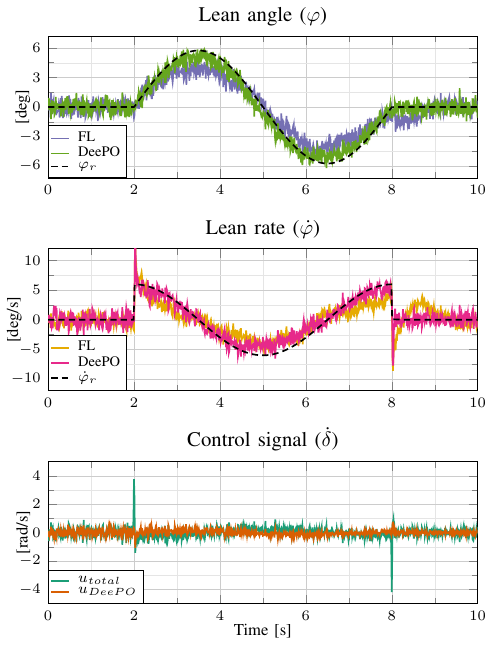}
    \caption{Tracking performance of the DeePO algorithm in simulation with $\xi = 1$. The top and middle plots illustrate the lean angle and rate tracking results for the DeePO+FL setup and the FL-only approach. The bottom plot presents the total control signal along with DeePO’s contribution.}
    \label{fig:simResults}
\end{figure}

The results for tracking a reference lean angle and lean rate are given in the top and middle plots of Fig.~\ref{fig:simResults}. The bottom plot represents the contribution of the DeePO algorithm to the system's total control signal. The update rate of the control gain is $\xi=1$, i.e., the control gain is updated with every sample, in Fig.~\ref{fig:simResults}. The results highlight the effectiveness of our unified control framework using FL and DeePO for tracking the reference lean angle, $\varphi\dti{t}{r}$, and lean rate, $\dot{\varphi}\dti{t}{r}$. The tracking is smoother, and the error is reduced compared to using only FL, as shown in the top and middle plots of Fig.~\ref{fig:simResults}. This improvement demonstrates DeePO's ability to adapt and refine the control policy iteratively, even under simulated nonlinear dynamics and sensor noise. The bottom plot of Fig.~\ref{fig:simResults} shows the contribution of DeePO to the total control signal and highlights that DeePO complements the FL controller and enables fine-tuning of the control signal online.

\begin{figure}[t]
    \centering
    \includegraphics{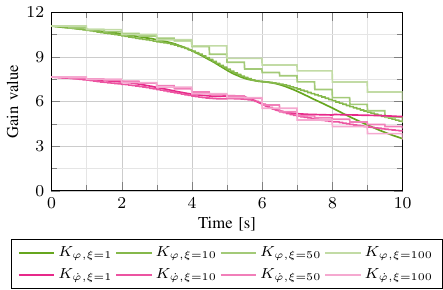}
    \caption{Evaluation of the control policy in simulation over time with different update rates of the control gain.}
    \label{fig:simControlPolicy}
\end{figure}

\begin{figure}[t]
    \centering
    \includegraphics{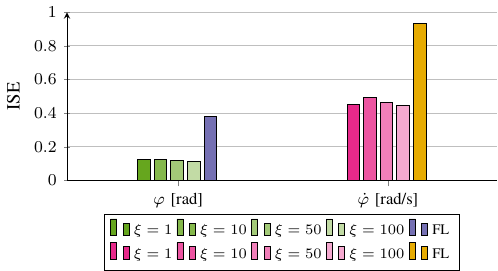}
    \caption{Integrated squared error of the lean angle and lean rate for different values of $\xi$ in simulation. The performance of only using FL is also included as a baseline.}
    \label{fig:simISE}
\end{figure}

In Fig.~\ref{fig:simControlPolicy}, the control policy for different update rates with respect to time is presented, and the corresponding ISE values are reported in Fig.~\ref{fig:simISE}, together with the ISE values when using only FL. These results clearly show that updating the control policy applied to the bicycle at every iteration of the algorithm is unnecessary. In fact, the performance improves for some of the lower update rates, i.e., higher values of $\xi$. However, it is not obvious how this parameter should be chosen, as for $\xi=10$, the performance is slightly worse than $\xi=1$. On the other hand, the performance is improved for $\xi=50$ and $\xi=100$. However, a low update rate may reduce DeePO's adaptation to fast changes in the system's behavior, reducing performance in systems with rapidly changing dynamics. Moreover, from Fig.~\ref{fig:simControlPolicy}, the control policy does not seem to converge, which may be due short simulation horizon. 

\begin{figure}[t]
    \centering
    \includegraphics{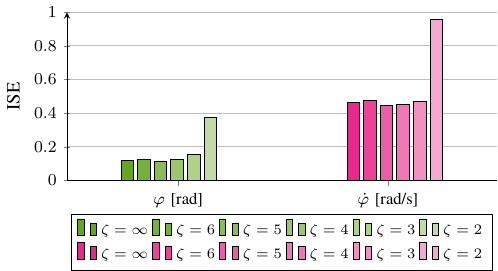}
    \caption{Integrated squared error of tracking the lean angle and lean rate references for different values of the forgetting factor in DeePO. The forgetting factor is defined as $\lambda = 1 - 10^{-\zeta}$.}
    \label{fig:simVarylambda}
\end{figure}

Finally, Fig.~\ref{fig:simVarylambda} presents the ISE values when the forgetting factor varies while the remaining control parameters are kept fixed. The results show that introducing a forgetting factor in DeePO may further enhance control performance. An intermediate forgetting factor of $\lambda=1-10^{-5}$ yields the best performance, while $\lambda = 1-10^{-2}$ results in the largest ISE value. Though varying $\xi$ and $\lambda$ may offer performance enhancements, finding the optimal and likely state-dependent values, for these parameters remains an area for future research. Moreover, the consistency of the simulation results highlights the efficiency of our unified control framework, which integrates FL and DeePO under varying circumstances.

\subsection{Experimental setup} 
\begin{figure}[t]
    \centering
    \includegraphics[width=\columnwidth]{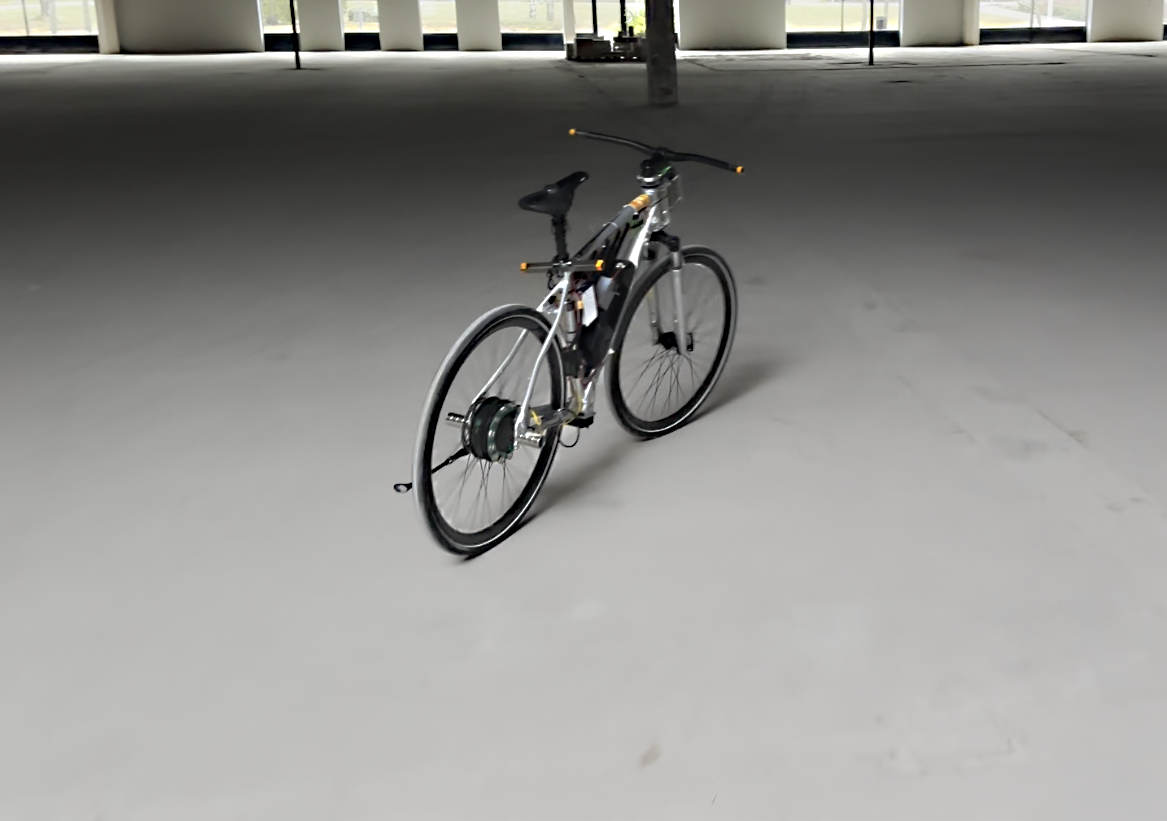}
    \caption{The indoor environment where the experiments were conducted on a flat concrete floor.}
    \label{fig:expBike}
\end{figure}

An initial experiment is conducted in which a dataset is collected from the sensors and the control signal. Similar to the initial simulation, the FL controller with the parameters from Table~\ref{tab:modParam} together with a persistently exciting input as $u_{PE} = \mathcal{N}(0,\sigma_{PE})$ are used as the input to the system, with $\sigma_PE=0.2$ rad/s. Furthermore, we consider tracking of a lean angle and lean rate reference of zero, and the tracking errors make up the states as in the simulation. The experiments are conducted in an indoor warehouse building with a flat concrete floor; see Fig.~\ref{fig:expBike}. The instrumented bicycle starts from a standstill, and the operator starts the bicycle using a switch on the RC controller and assists in balancing the bicycle until it reaches its constant goal velocity of $8$ km/h. Using a second switch on the RC controller, the experiment is initiated, and the persistently exciting input, together with the FL controller, regulates the steering velocity of the handlebar and, by extension, the bicycle's balance. The Xsens MTi-7 and the Dynamixel data are collected at a sampling frequency of $100$ Hz on the Raspberry Pi. Algorithm~\ref{alg:deepo} and the FL controller, implemented on the Raspberry Pi, compute a new control signal every $0.01$ s and transmit it to the Dynamixel servo. 

The sampled input and state data are post-processed in Matlab, where they are used to construct $U_{0,T}, X_{0,T}, X_{1,T}$, with $T=300$. Next, the initial policy is obtained by solving~\eqref{prob:regu} in Matlab, with $Q=I_2$, $R=0.01$, and $\gamma=1$. In the subsequent experiments, we set the forgetting factor and learning rate as $\lambda = 1-10^{-4}$ and $\eta=10^{-3}$, respectively. While $\gamma$, $\lambda$, and $\eta$ are kept the same in the experiments as in the simulation, the $Q$, $R$, and $T$ are changed in experiments. The increase in the number of data samples can be explained by the idealized dynamics in simulation compared to experiments where, for instance, mechanical imperfections, delays, and external disturbances are present. Thus, a larger dataset ensures a more reliable estimation of the system's behavior. Moreover, the increase in the weight of $\dot\varphi$ in $Q$ is justified by the sensor noise and the need for robustness in experiments. Finally, in simulations, the resolution of the  steering motor is neglected, and the complex dynamics of the actuating steering system are not fully captured. Thus, we can allow for more aggressive control actions using a lower value of $R$. On the other hand, in experiments, the resolution of the motor limits the possible steering commands. A larger $R$ in experiments ensures a smoother, more physically feasible control input. The same time-varying reference lean angle and lean rate used in simulations are also utilized in the experiments. Furthermore, we conduct several experiments where $\xi$ varies as $\xi=1, 10, 50,$ and $100$. Additionally, one experiment is conducted with only the FL controller as a baseline.

\subsection{Experimental results}
\begin{figure}[t]
    \centering
    \includegraphics{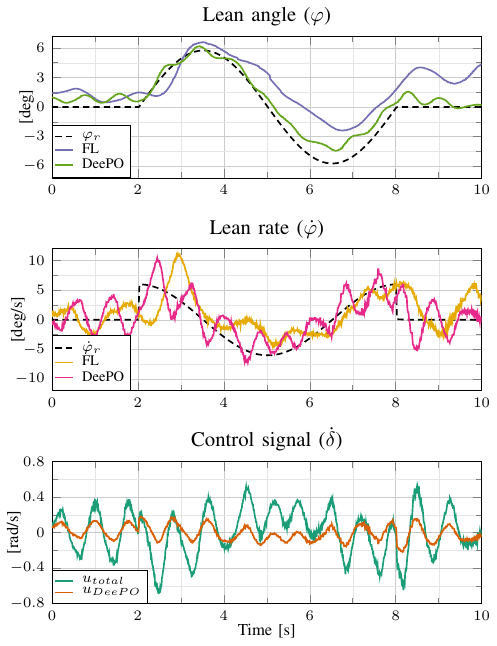}
    \caption{Experimental results of DeePO where the control policy is updated at every time step, i.e., $\xi=1$.}
    \label{fig:expResults}
\end{figure}

The lean angle and lean rate tracking performance of the DeePO algorithm is illustrated in the top and middle plots of Fig.~\ref{fig:expResults} using $\xi = 1$. The results of using only FL are also included in the figure. The control signal of DeePO and the total control signal of the system are highlighted in the bottom plot of Fig.~\ref{fig:expResults}. The results demonstrate the effectiveness of DeePO and its robust tracking of the reference lean angle and lean rate, with noticeable improvements compared to the FL controller, as evident from the first two plots of Fig.~\ref{fig:expResults}. The results also show how DeePO adapts over time, even in the presence of nonlinearities, sensor noise, and external disturbances (e.g., floor imperfections or variations in tire grip). However, the oscillations in lean angle and lean rate have a much higher amplitude in experiments compared to simulations, which the steering motor's resolution limitation could partly explain. In experiments, the resolution is limited to $0.024$ rad/s per unit, a factor not considered in simulations. Moreover, simulations have unmodelled dynamics and environmental details compared to experiments, such as joint friction, uneven terrain, and approximations made in the model. 

\begin{figure}[t]
    \centering
    \includegraphics{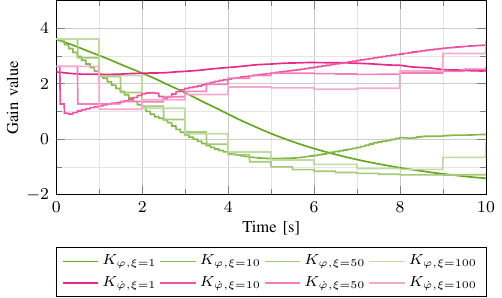}
    \caption{Evolution of control gains in experiments for different values of $\xi$.}
    \label{fig:expControlPolicy}
\end{figure}

\begin{figure}[t]
    \centering
    \includegraphics{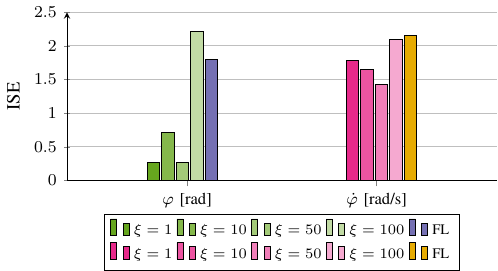}
    \caption{Integrated squared error for different values of $\xi$ and when using the FL controller alone in experiments.}
    \label{fig:expISE}
\end{figure}

The evolution of the control gain values for the lean angle and lean rate are reported in Fig.~\ref{fig:expControlPolicy} with varying values for gain update frequency $\xi$.  As observed in simulations, the evolution of the control gains in experiments follows the same trend,  even though the update frequencies vary. However, the difference between the evolution of the control gains in experiments and the control gains of the simulation, as presented in Fig.~\ref{fig:simControlPolicy}, is quite different, which indicates a gap between the simulations and experiments. It also highlights the usefulness of adaptive control methods, which can refine the feedback gain based on online experiment data. The gap between simulations and experiments is also evident when comparing the ISE values in Fig.~\ref{fig:simISE} and Fig.~\ref{fig:expISE}. In experiments, update at every time step or intermediate rates of $\xi=50$ produces significantly better results than update at lower rates or using only FL. The considerably higher ISE for $\xi=100$ and using only FL further highlights the limitations of static or infrequent control gain updates. In simulations where we have control over the initial conditions, noise, and external disturbances, the results are more aligned, and the impact of $\xi$ is not as evident as in experiments. One particular challenge in the experiments was finding a suitable pre-stablizing control form initial data and control of the initial conditions. A video demonstration of the simulations and experiments is available online~\footnote{\href{https://youtu.be/5RKnr6tPiuw}{https://youtu.be/5RKnr6tPiuwonline}}.

\section{Conclusion}
\label{sec:5}
This paper introduced a unified framework that balances an autonomous bicycle by combining an FL controller in the inner loop and DeePO in the outer loop. The primary objective of the FL loop is to stabilize and partially linearize an otherwise unstable and nonlinear system. However, practical systems often contain unmodeled dynamics and time-varying characteristics that can degrade the performance of the FL controller when used in isolation. To address these challenges, we integrated a DeePO controller on top of the FL loop.

We derived an initial control policy using a finite set of offline, persistently exciting input and state data. To handle the nonlinearities and disturbances that may degrade the policy's performance, we introduced a robustness-promoting regularizer to refine the initial stabilizing policy and a forgetting factor in the DeePO framework to adapt to the time-varying nature of our case study.

We demonstrated the effectiveness of the DeePO+FL approach through both simulations and real-world experiments on an autonomous bicycle. The results clearly showed that DeePO+FL outperforms the FL-only approach, particularly in terms of tracking the reference lean angle and lean rate more accurately. Additionally, we evaluated the impact of the control gain update frequency and found that performance improvements could be achieved with a lower update frequency. However, determining the optimal update rate remains an open question for future research.

The experimental and simulation results demonstrated that the proposed controller effectively adapts to the system dynamics despite the presence of nonlinearities, sensor noise, and hardware limitations. Our work illustrates the potential of direct data-driven methods to adapt and control nonlinear systems, such as an autonomous bicycle, relying solely on data. In the future, we plan to enhance the DeePO algorithm by incorporating the robustness regularizer in its online component. Additionally, exploring direct data-driven navigation for the bicycle is another exciting direction for further research.

\bibliographystyle{IEEEtran}  
\bibliography{refs.bib}
\def\spbio{30}
\vspace{-\spbio pt}
\begin{IEEEbiography}[{\includegraphics[width=1in,height=1.25in,clip,keepaspectratio]{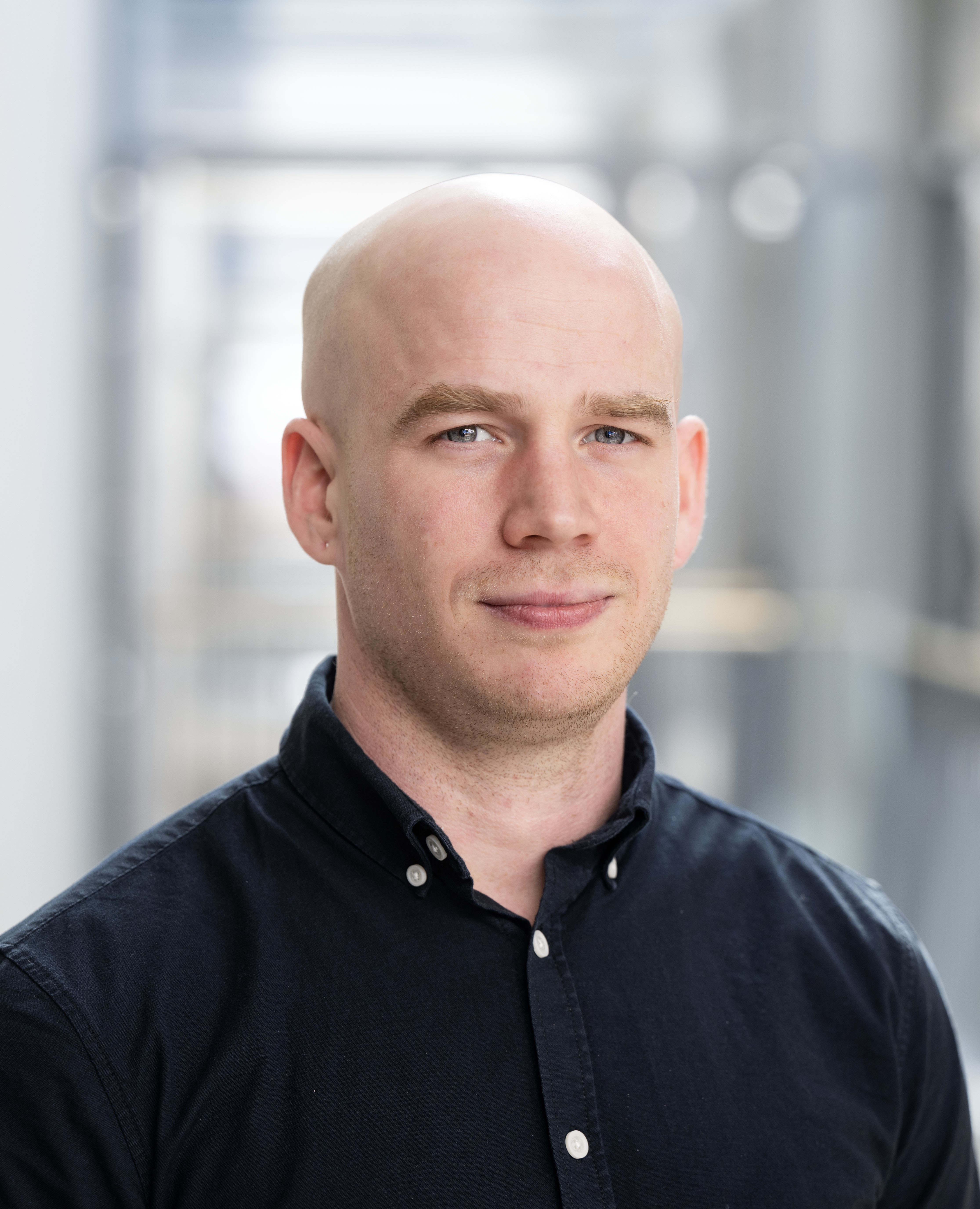}}] {Niklas Persson} received an M.Sc in Robotics from M{\"a}lardalen University, V{\"a}ster{\aa}s, Sweden in 2019. Since 2020, he has been pursuing a PhD degree in electronics at the Intelligent Future Technologies division of M{\"a}lardalen University, working on the control and navigation of autonomous bicycles. In 2023, he received a Licentiate degree at Mälardalen University. His research interests include autonomous robots and vehicles, control theory, and embedded systems. 
\end{IEEEbiography}
\vspace{-\spbio pt}

\begin{IEEEbiography}[{\includegraphics[width=1in,height=1.25in,clip,keepaspectratio]{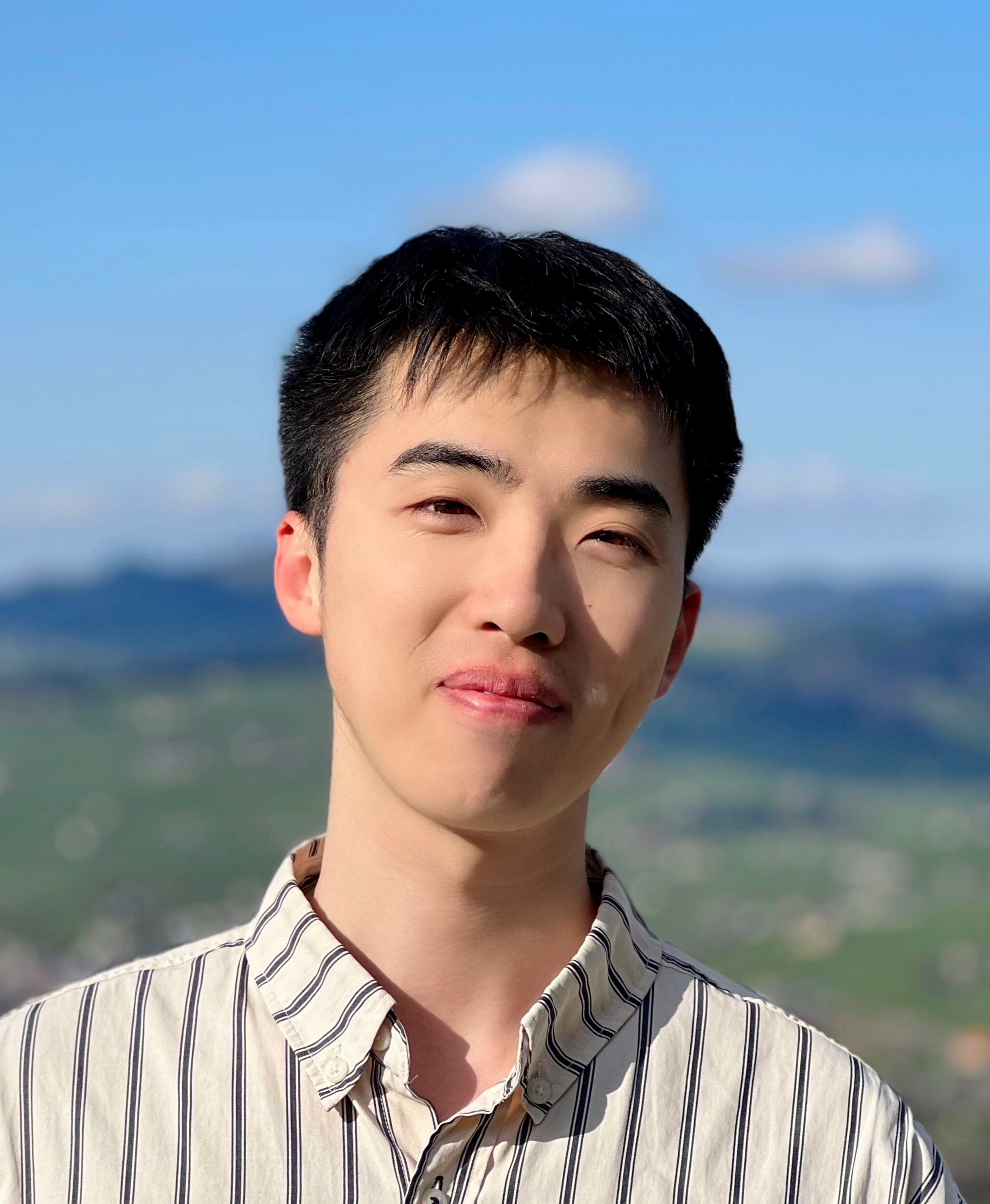}}] {Feiran Zhao} received the B.S. degree in Control Science and Engineering from the Harbin Institute of Technology, China, in 2018, and the Ph.D. degree in Control Science and Engineering from the Tsinghua University, China, in 2024. He is now a postdoc at ETH Z\"{u}rich. His research interests include policy optimization, data-driven control, adaptive control and their applications. 
\end{IEEEbiography}
\vspace{-\spbio pt}

\begin{IEEEbiography}[{\includegraphics[width=1in,height=1.25in,clip,keepaspectratio]{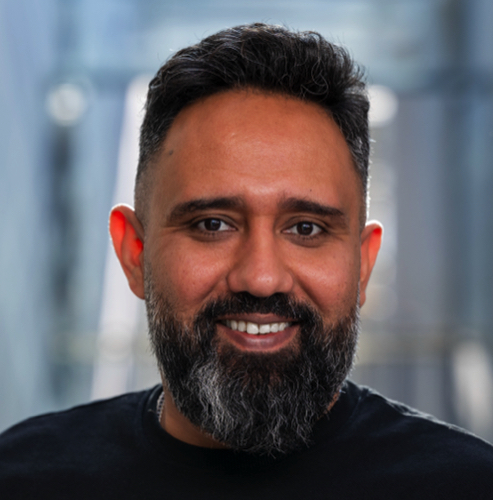}}] {Mojtaba Kaheni} (SM'25) is a Postdoctoral Researcher at the {School of Innovation, Design, and Technology (IDT)}, {Mälardalen University}, Västerås, Sweden. He received his {M.Sc.} and {Ph.D.} in Control Engineering from {Shahrood University of Technology}, Shahrood, Iran, in 2011 and 2019, respectively.
Dr. Kaheni has held visiting scholar positions at the {University of Florence}, Italy, and {Lund University}, Sweden. From August 2020 to December 2022, he served as a Postdoctoral Researcher at the {University of Cagliari}, Italy.
His research interests include {control theory}, {distributed optimization}, {multi-agent systems}, and {resiliency}.

\end{IEEEbiography}
\vspace{-\spbio pt}

\begin{IEEEbiography}
[{\includegraphics[width=1in,height=1.25in,clip,keepaspectratio]{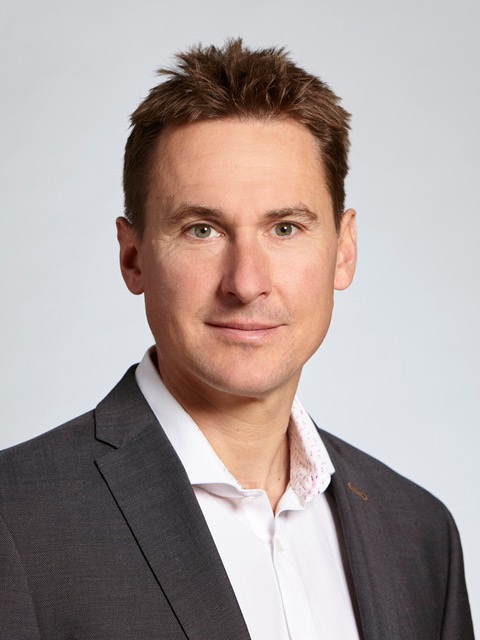}}]	
	{Florian D\"{o}rfler} is a Full Professor at the Automatic Control Laboratory at ETH Z\"{u}rich. He received his Ph.D. degree in Mechanical Engineering from the University of California at Santa Barbara in 2013, and a Diplom degree in Engineering Cybernetics from the University of Stuttgart in 2008. From 2013 to 2014 he was an Assistant Professor at the University of California Los Angeles. He has been serving as the Associate Head of the ETH Z\"{u}rich Department of Information Technology and Electrical Engineering from 2021 until 2022. His research interests are centered around automatic control, system theory, and optimization. His particular foci are on network systems, data-driven settings, and applications to power systems. He is a recipient of the distinguished young research awards by IFAC (Manfred Thoma Medal 2020) and EUCA (European Control Award 2020). His students were winners or finalists for Best Student Paper awards at the European Control Conference (2013, 2019), the American Control Conference (2016, 2024), the Conference on Decision and Control (2020), the PES General Meeting (2020), the PES PowerTech Conference (2017), the International Conference on Intelligent Transportation Systems (2021), and the IEEE CSS Swiss Chapter Young Author Best Journal Paper Award (2022, 2024). He is furthermore a recipient of the 2010 ACC Student Best Paper Award, the 2011 O. Hugo Schuck Best Paper Award, the 2012-2014 Automatica Best Paper Award, the 2016 IEEE Circuits and Systems Guillemin-Cauer Best Paper Award, the 2022 IEEE Transactions on Power Electronics Prize Paper Award, and the 2015 UCSB ME Best PhD award. He is currently serving on the council of the
	European Control Association and as a senior editor of Automatica.
	\end{IEEEbiography}
\vspace{-\spbio pt}
\begin{IEEEbiography}[{\includegraphics[width=1in,height=1.25in,clip,keepaspectratio]{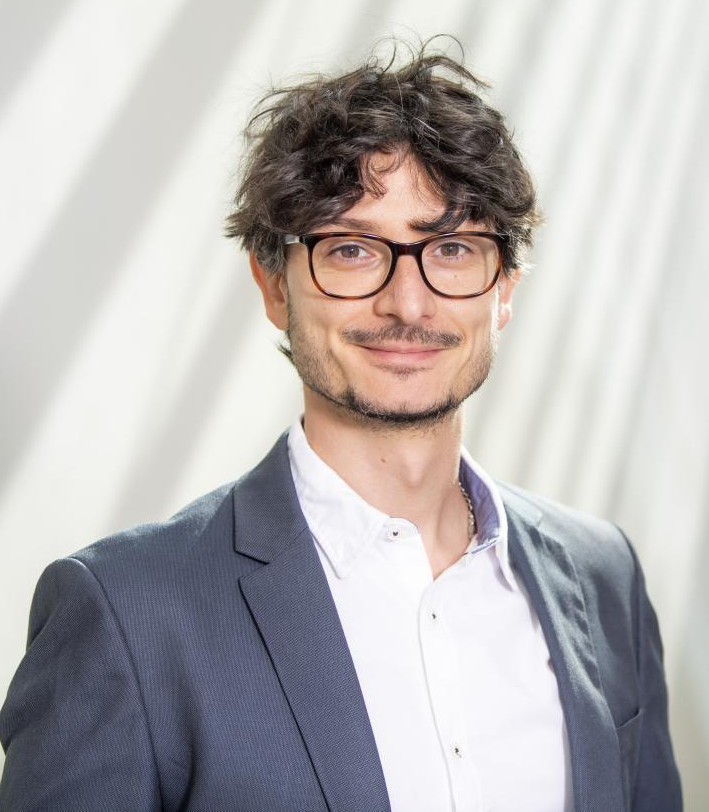}}]{Alessandro~V. Papadopoulos} (SM'19)
is a Full Professor of Electrical and Computer Engineering at M{\"a}lardalen University, V{\"a}ster{\aa}s, Sweden, and a QUALIFICA Fellow at the University of M{\'a}laga, Spain. Since March 2024, he has been the scientific leader of Applied AI at M{\"a}lardalen University. He received his B.Sc. and M.Sc. (summa cum laude) degrees in Computer Engineering from the Politecnico di Milano, Milan, Italy, and his Ph.D. (Hons.) degree in Information Technology from the Politecnico di Milano, in 2013. He was a Postdoctoral researcher at the Department of Automatic Control, Lund, Sweden (2014-2016) and Politecnico di Milano, Milan, Italy (2016). 
He was the Program Chair for the Mediterranean Control Conference (MED) 2022, the Euromicro Conference on Real-Time Systems (ECRTS) 2023, and the ACM/SPEC International Conference on Performance Engineering (ICPE) 2025. He is an associate editor for the ACM Transactions on Autonomous and Adaptive Systems, Control Engineering Practice, and Leibniz Transactions on Embedded Systems.
His research interests include robotics, control theory, real-time systems, and autonomic computing. 
\end{IEEEbiography}

\end{document}